\documentclass[12pt,journal,draftclsnofoot,onecolumn,letterpaper]{IEEEtran} 

\usepackage{amssymb}
\usepackage{amsmath}
\usepackage{amsmath,bm}
\usepackage{amsthm}
\usepackage{graphicx}
\usepackage{subfigure}
\usepackage{cite}
\usepackage{enumerate}
\usepackage[ruled]{algorithm2e}
\usepackage{color, soul}

\begin{document}

\newtheorem{definition}{\bf ~~Definition}
\newtheorem{observation}{\bf ~~Observation}
\newtheorem{lemma}{\bf ~~Lemma}
\newtheorem{proposition}{\bf ~~Proposition}
\newtheorem{remark}{\bf ~~Remark}

\title{D2D-U: Device-to-Device Communications in Unlicensed Bands for 5G and Beyond}
\author{
\IEEEauthorblockN{
{Hongliang Zhang}, \IEEEmembership{Student Member, IEEE},
{Yun Liao}, \IEEEmembership{Student Member, IEEE},
{and Lingyang Song}, \IEEEmembership{Senior Member, IEEE}}\\

\vspace{-0.5cm}

\thanks{H. Zhang, Y. Liao, and L. Song are with School of Electronics Engineering and Computer Science, Peking University, Beijing, China (email: \{hongliang.zhang,yun.liao,lingyang.song\}@pku.edu.cn).}
\thanks{Part of this work has been accepted by Proc. IEEE ICC 2017 \cite{HYL-2017}.}
}
\maketitle

\vspace{-0.5cm}

\begin{abstract}

Device-to-Device (D2D) communication, which enables direct communication between nearby mobile devices, is an attractive add-on component to improve spectrum efficiency and user experience by reusing licensed cellular spectrum in 5G system. In this paper, we propose to enable D2D communication in unlicensed spectrum~(D2D-U) as an underlay of the uplink LTE network for further booming the network capacity. A sensing-based protocol is designed to support the unlicensed channel access for both LTE and D2D users. We further investigate the subchannel allocation problem to maximize the sum rate of LTE and D2D users while taking into account their interference to the existing Wi-Fi systems. Specifically, we formulate the subchannel allocation as a many-to-many matching problem with externalities, and develop an iterative user-subchannel swap algorithm. Analytical and simulation results show that the proposed D2D-U scheme can significantly improve the system sum-rate.
\end{abstract}


\newpage

\section{Introduction}%
\label{Introduction}

With the explosive growth of mobile devices and bandwidth-hungry applications such as video streaming and multimedia file sharing, user demands for mobile broadband are undergoing an unprecedented rise, which pushes the limits of current 4G LTE systems \cite{KMCCK-2009}. To improve spectrum efficiency and user experience, device-to-device~(D2D) communications underlaying LTE networks have been proposed as a promising approach to facilitate high data rate services in a short range and boost the performance of LTE systems~\cite{KMCCK-2009,GEGSNGZ-2012,LZCX-2012} for future 5G communications and beyond. D2D communications enable mobile devices in proximity to establish a direct link without traversing the base station~(BS), and reuse the spectrum with the LTE system by the control of the BS, which enjoy the benefits of fast access to the radio spectrum in terms of proximity gain, reuse gain, and paring gain~\cite{LDZE-2015,CKCO-2011,HJSD-2011,HLZ-2016}. 

Recently, the operators expand LTE services to unlicensed spectrum to alleviate congestion. Mobile traffic offloading is a conventional method to utilize the unlicensed spectrum, in which the data is offloaded to Wi-Fi networks~\cite{AHM-2013,KJYIB-2013,MMAWSM-2013,HPD-2012}. However, the offloading schemes commonly suffer from low efficiency and poor guarantee of quality-of-service~(QoS) due to the inferior performance of Wi-Fi and the lack of coordination between cellular and Wi-Fi systems \cite{YYLMLZ-2015}. In light of these issues, the 3rd Generation Partnership Project~(3GPP) has initialed the research on licensed assisted access~(LAA) to integrate the unlicensed carriers with the licensed ones for data transmission~\cite{3GPP-2013}. Based on the LAA scheme, the LTE-unlicensed~(LTE-U) technology is proposed to extend LTE to the unlicensed spectrum by the existing carrier aggregation~(CA) technology~\cite{ASQE-2015,RMLZXL-2015,QGHAGA-2016,GGLAJD-2016,RGAG-2016,RGAG-2016-2}.

As LTE-U technology shows satisfying performance, D2D communications underlaying LTE networks in unlicensed spectrum becomes a natural solution to further improve system throughput, in particular hotspot areas with large number of D2D links. However, due to the mutual interference among LTE-U network, D2D users, and the opportunistic feature of unlicensed channel access in existing Wi-Fi systems, D2D-Unlicensed~(D2D-U) communication turns out to be much complicated. In this paper, we investigate the underlaid D2D communications in unlicensed spectrum. Note that different from most previous peer-to-peer communication technologies in unlicensed spectrum such as Wi-Fi Direct~\cite{YHL-2015,DAKVP-2014,WiFi-Direct}, which builds the network upon the IEEE 802.11 infrastructure mode and allows users to negotiate with each other in an AP-like method, D2D-U requires assist and control from the central BS. With the involvement of BS, D2D users can work as an underlay of LTE system in both licensed and unlicensed spectra.

As aforementioned, the major challenges of implementing D2D-U are (1) the opportunistic feature of unlicensed channel access due to current 802.11 mechanism adopted by Wi-Fi systems; and (2) the interference management issue among the three types of systems, i.e., the access and transmission of D2D-U users do not cause significant interference to the existing Wi-Fi system as well as the LTE-U system. To cope with the first challenge and be compatible with current LTE standards\footnote{The work in \cite{YWHLSXJ-2016} proposed an access protocol based on listen-before-talk~(LBT) mechanism to mitigate collision with the ongoing Wi-Fi transmissions. However, LBT requires changes to the LTE specifications.}, we design a duty cycle based protocol \cite{Qualcomm-2014,CD-2016,ATKHM-2015}, in which the BS schedules transmissions according to the data demand. To tackle the second challenge, unlike the work in~\cite{RGFZ-2016} which only maximizes the total sum-rate, we investigate the subchannel allocation problem to leverage the maximization of the sum-rate of LTE-U and D2D-U users and the protection of Wi-Fi performance. This subchannel allocation problem is originally a mixed-integer non-linear programming (MINLP) problem, which is generally NP-hard. For this reason, we reformulate it as a many-to-many game with externalities~\cite{HBYNLXQ-2014, BLY-2016, OWSMH-2015, ECABA-2011, AM-1992}, and solve it with low computational complexity by designing an iterative user-subchannel swap-matching algorithm.


The major contributions of this paper are summarized as follows.
\begin{itemize}
\item We propose a feasible duty cycle based protocol for the LTE-U and D2D-U users to utilize the unlicensed spectrum.
\item An approximated model is elaborated to evaluate the interference to Wi-Fi networks introduced by LTE-U and D2D-U users.
\item We investigate the subchannel allocation problem by a many-to-many matching game with externality, and analyze its stability, convergence, complexity, and optimality.
\end{itemize}

The rest of the paper is organized as follows. In Section \ref{System}, we first introduce the system model for the coexistence among LTE, D2D, and Wi-Fi users, as well as their PHY/MAC features, and then discuss the interference issues. In Section \ref{protocoldesign}, a duty cycle based protocol is elaborated to support LTE-U and D2D-U users accessing the unlicensed spectrum. Then we formulate the optimization problem for subchannel allocation as a many-to-many matching game with externalities in Section \ref{problem}. In Section \ref{many-to-many}, we develop an iterative algorithm to solve the many-to-many matching game with its property analysis. In Section \ref{Discussion}, the system performance is discussed. Numerical results in Section \ref{Simulation} evaluate the proposed algorithm and the performance of the D2D-U. Finally, conclusion remarks are drawn in Section \ref{Conclusion}.

\section{System Model}%
\label{System}

In this section, we present the coexistence scenario of LTE, D2D, and Wi-Fi systems in both licensed and unlicensed spectra. Then, the characteristics of LTE, D2D, and Wi-Fi systems in the MAC and PHY layers are elaborated respectively. Furthermore, we discuss the interference issue within the coexistence network at the end.

\subsection{Scenario Description}

\begin{figure}[!t]
\centering
\includegraphics[width=3.5in]{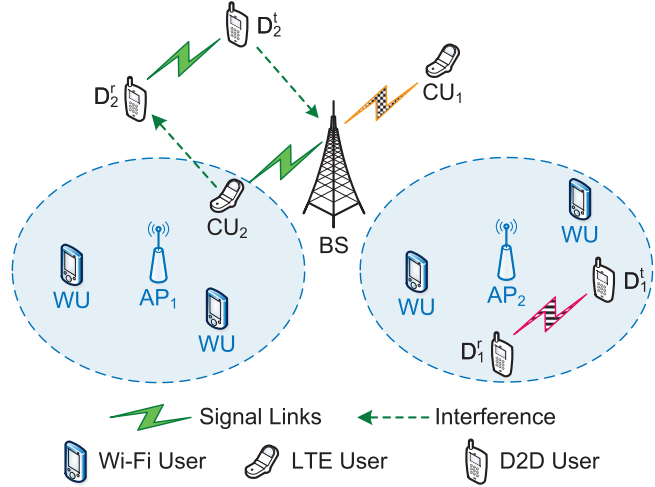}
\vspace{-5mm}
\caption{System model for LTE, D2D, and Wi-Fi coexistence in both licensed and unlicensed spectra.}
\vspace{-7mm}
\label{scenario}
\end{figure}

As shown in Fig.~\ref{scenario}, we consider an uplink scenario in an LTE network with one BS and $Q$ Wi-Fi access points (APs), denoted by $\mathcal{Q} = \{1,\ldots,Q\}$. There exist $N$ LTE users, denoted by CU$_n~(n \in \mathcal{N} = \left\{1,\ldots,N\right\})$, and $M$ D2D users, denoted by $\left( \text{D}_m^t, \text{D}_m^r \right)~(m \in \mathcal{M} = \{1,\ldots, M\})$, where D${_m^t}$ and D${_m^r}$ represent the transmitter and receiver of D2D user D$_m$, respectively. The system owns $K$ licensed subchannels with uniform bandwidth $B_l$ to support orthogonal frequency division multiple access~(OFDMA) transmissions, denoted by $\mathcal{K} = \{1,\ldots,K\}$.

For the Wi-Fi system, we assume that within the coverage of AP$_q~(q \in \mathcal{Q})$, there exist $F_q$ Wi-Fi users marked by WU$_f^q~(f = 1, \ldots, F_q)$. Besides, we assume that there are $L$ unlicensed channels to support different APs, e.g., there are 23 channels for IEEE 802.11n in the 5GHz band, and BS will select one of them to support LTE-U and D2D-U users. Since the bandwidth of a unlicensed channel is much wider than one licensed subchannel in LTE system, each LTE or D2D user only requires a fraction of the unlicensed channel. Thus, to reuse the unlicensed channel more efficiently, the unlicensed channel is divided to $K^u$ unlicensed subchannels with bandwidth $B_u$~\cite{JXLZLD-2015}, marked by $\mathcal{K}^u = \{K+1,\ldots,K+K^u\}$, so that multiple LTE-U users and D2D pairs can transmit on the unlicensed channel concurrently.

\subsection{Characteristics of LTE-U, D2D, and Wi-Fi Networks}

In this part, we sequentially elaborate the PHY/MAC characteristics of the coexisting systems, i.e., LTE and D2D users in the licensed/unlicensed bands, and existing Wi-Fi characteristics in the unlicensed band.

We assume that all devices transmit with fixed power in this work. Specifically, the transmit power of an LTE and D2D transmitter on any subchannel is fixed on $P^c$ and $P^d$, respectively; and the transmit power of the APs as well as the Wi-Fi users over the whole unlicensed channel is fixed on $P^w$.
The free space propagation path-loss model with Rayleigh fading is adopted to model the channel gain between any two devices in the network, i.e., for the link from device $i$ to device $j$, the received power can be expressed as
\begin{equation}\label{Received_power}
p_{i,j}^r = p^t_i \cdot |h_{i,j}|^2 \xi_{i,j} = p^t_i \cdot G \cdot d_{i,j}^{-\alpha} \cdot |h_0|^2 \xi_{i,j},
\end{equation}
where $p^t_i$ represents the transmit power of user $i$, $d_{i,j}$ is the distance between devices $i$ and $j$, $\alpha$ is the pathloss exponent, $G$ is the constant power gains factor introduced by amplifier and antenna, $h_0 \sim \mathcal{CN}(0,1)$ is a complex Gaussian variable representing Rayleigh fading, and $\xi_{i,j}$ follows log-normal distribution representing shadowing fading. Besides, we assume that the thermal noise at each device satisfies independent Gaussian distribution with zero mean and the same variance $\sigma ^2$.

\subsubsection{LTE network and underlaid D2D users}

In PHY layer, LTE and D2D users can utilize both the licensed and unlicensed spectra orthogonally. We assume that each user is able to occupy multiple subchannels. In addition, to guarantee reliable transmission of the control signaling, an active LTE/D2D user must hold at least one licensed subchannel \cite{RMLZXL-2015}. Similar to the channel sharing in the licensed spectrum, D2D-U users can work as an underlay of the LTE-U users. In other words, D2D users can utilize the licensed/unlicensed subchannels concurrently occupied by some LTE users.

As for the MAC layer, the LTE/D2D systems adopt a centralized MAC protocol. The BS controls the access of both types of users and decides the subchannel allocation in a centralized manner to mitigate mutual interference or maximize the system sum-rate.

\subsubsection{Wi-Fi systems}

The Wi-Fi systems operate only in the unlicensed spectrum. Different from the OFDMA-based channel utilization in LTE systems, the Wi-Fi transmission covers the whole unlicensed channel. Thus, Wi-Fi systems only allow one user to occupy the channel at a time.

For the MAC layer, without a central controller, the Wi-Fi systems adopt a sensing and contention-based MAC protocol, i.e., carrier sense multiple access with collision avoidance (CSMA/CA) \cite{KWY-2014}. Specifically, before transmission, a Wi-Fi user first listens to the intended channel. If the channel is unoccupied, the Wi-Fi user begins backoff process to avoid collision. Otherwise, the Wi-Fi user keeps sensing until the channel is judged idle.

\subsection{Evaluation of Interference to Wi-Fi Systems}

When LTE-U and D2D-U users occupy the unlicensed channel, the nearby Wi-Fi users cannot access, and thus the performance of Wi-Fi system would be severely degraded. To quantify the performance degradation brought by LTE-U and D2D-U users, we introduce the definition of \emph{interference range} on the Wi-Fi network. Within the interference range, each Wi-Fi user is able to detect the channel unavailable, and then suspend their transmission attempts. In practice, the fading value varies between subframes, and thus, it is difficult for BS to detect the interference range of each LTE-U or D2D-U user at the beginning of a subframe.  To better model the interference, the interference range of a LTE-U/D2D-U user is defined as the area where the expectation of the received power from this user exceeds the power threshold. Therefore, the interference range to the Wi-Fi network is a circle centered at the transmitter, whose radius is positively related to the transmit power. Intuitively, the users with large interference range has low probability of utilizing unlicensed subchannels, because large numbers of Wi-Fi users will be interfered by this user. On contrary, the users with small interference range are more likely to utilize unlicensed subchannels due to their limited interference to the Wi-Fi network.

However, when multiple LTE-U and D2D-U users transmit on the unlicensed spectrum concurrently, their individual interference circles may overlap, which is hard to derive the closed form expression for the area of the total interference range. For this reason, in the following of this part, we present an approximated model of the interference range to evaluate the performance degradation in Wi-Fi system. Intuitively speaking, a smaller interference range can be obtained if the BS allocates the unlicensed subchannels to those adjacent users rather than those whose interference ranges do not overlap. Inspired by this observation, we use the minimum distance between a new LTE-U/D2D-U user to others to approximate the additional interference range introduced by this add-on user. Let $L_c$ and $L_d$ denote the radii of individual interference circles of LTE-U and D2D-U users, respectively, where $L_d \leq L_c$. And we define $\mathcal{C}^u$ as the user set in which users utilize unlicensed subchannels. With these notations, the weight functions for LTE-U and D2D-U users are given as below.

\begin{figure}[!t]
\centering
\includegraphics[width=3.5in]{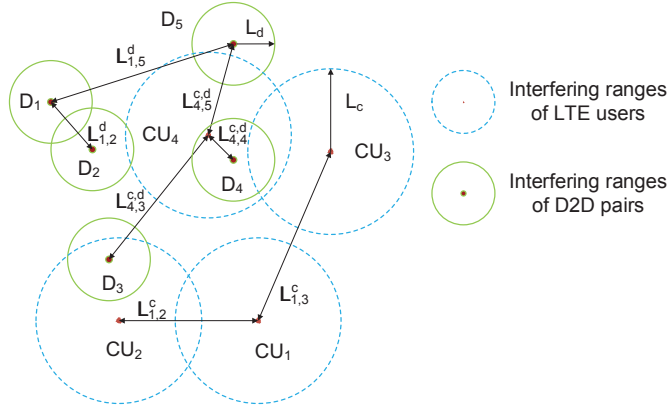}
\vspace{-5mm}
\caption{Illustrations for interference range.}
\vspace{-7mm}
\label{illustration}
\end{figure}

\subsubsection{LTE-U user}
When LTE-U user CU$_n$ is allocated to the unlicensed subchannel the first time, and CU$_j$ is an LTE-U user, the increased area of interference range is related to the distance between CU$_n$ and CU$_j$, denoted by $L^{c}_{n,j}$\footnote{LTE users CU$_n$ and CU$_j$ will inform the BS of their locations, and then the BS can calculate the distance between these two users.}. When the interference ranges of these two LTE users overlaps, i.e., $L^{c}_{n,j} < 2L_c$, as CU$_1$ and CU$_2$ in Fig.~\ref{illustration}, we assume that the weight is proportional to the distance $L^{c}_{n,j}$. As CU$_1$ and CU$_3$ illustrated in Fig.~\ref{illustration}, their interference ranges do not overlap, i.e., $L^{c}_{n,j} \geq 2L_c$, the increased area will not grow as the distance $L^{c}_{n,j}$. Therefore, the weight function for the increased area $f^c_{n,j}$ can be expressed as
\begin{equation}\label{function1}
f^{c}_{n,j} =
\left\{\begin{array}{ll}
L^{c}_{n,j},~~&L^{c}_{n,j} < 2L_c,\hfill\\
2L_c,~~&L^{c}_{n,j} \geq 2L_c.\hfill\\
\end{array} \right.
\end{equation}

On the other hand, given D2D-U user D$_m$, the increased area of interference range is also related to the distance between CU$_n$ and D$_m$, $L^{c,d}_{n,m}$. Note that $L_d \leq L_c$, when the interference range of D$_m$ is contained by that of CU$_n$ as CU$_4$ and D$_4$ in Fig.~\ref{illustration}, i.e., $L^{c,d}_{n,m} \leq L_c - L_d$, the increased area is proportional to the increased diameter $2(L_c - L_d)$. When the interference range of CU$_n$ overlaps with but does not contain that of D$_m$ as D$_5$ and CU$_4$, that is, $L_c - L_d < L^{c,d}_{n,m} < L_c + L_d$, the weight is proportional to the distance $L^{c,d}_{n,m}$ as well. Besides, when the interference ranges of D$_m$ and CU$_n$ do not overlap as CU$_4$ and D$_3$, i.e., $L^{c,d}_{n,m} \geq L_c + L_d$, the increased area is a constant. Therefore, the weight function for the increased area $f^{c,d}_{n,m}$ is written by
\begin{equation}\label{function2}
f^{c,d}_{n,m} =
\left\{\begin{array}{ll}
2(L_c - L_d),~~&L^{c,d}_{n,m} \leq L_c - L_d,\hfill\\
L^{c,d}_{n,m} + L_c - L_d,~~&L_c - L_d < L^{c,d}_{n,m} < L_c + L_d,\hfill\\
2L_c,~~&L^{c,d}_{n,m} \geq L_c + L_d.\hfill\\
\end{array} \right.
\end{equation}

The weight of LTE-U user CU$_n$ is the minimum increased interference range between CU$_n$ and any user allocated to unlicensed subchannels., that is,
\begin{equation}\label{function3}
f^{c}_{n} = \min\limits_{j,m \in \mathcal{C}^u}(f^c_{n,j},f^{c,d}_{n,m}).
\end{equation}

\subsubsection{D2D-U user}
Similar to the LTE-U users, if D2D-U user D$_m$ is the first time to utilize unlicensed subchannels, and D$_j$ is a D2D-U user, the increased range is related to the distance between D$^t_m$ and D$^t_j$. As illustrated in Fig.~\ref{illustration}, the increased range can also be calculated under two conditions: (1) the interference ranges of these two D2D users overlap; (2) their interference ranges do not overlap. Thus, the weight function $f^d_{m,j}$ between D$_m$ and D$_j$ is provided by
\begin{equation}\label{function4}
f^d_{m,j} =
\left\{\begin{gathered}
L^{d}_{m,j},~~L^{d}_{m,j} < 2L_d,\hfill\\
2L_d,~~L^{d}_{m,j} \geq 2L_d.\hfill\\
\end{gathered} \right.
\end{equation}

In addition, if there already exists LTE-U user CU$_n$, the increased area $f^{c,d}_{n,m}$ is also related to the distance $L^{d,c}_{m,n}$ between CU$_n$ and D$_m$, which can be given by
\begin{equation}\label{function5}
f^{d,c}_{n,m} =
\left\{\begin{array}{ll}
0,~~&L^{d,c}_{m,n} \leq L_c - L_d,\hfill\\
L^{c,d}_{n,m} + L_d - L_c,~~&L_c - L_d < L^{d,c}_{m,n} < L_c + L_d,\hfill\\
2L_d,~~&L^{d,c}_{m,n} \geq L_c + L_d.\hfill\\
\end{array} \right.
\end{equation}
Therefore, the weight of D$_m$ is
\begin{equation}\label{function6}
f^{d}_{m} = \min\limits_{j,n \in \mathcal{C}^u}(f^d_{m,j},f^{d,c}_{m,n}).
\end{equation}

\subsection{Interference in the LTE-U/D2D network}

The mutual interference between the LTE and D2D users is analyzed in this subsection. We assume that one subchannel can be allocated to a maximum of one LTE user, and a subchannel can be allocated to at most $V^s$ users for the sake of QoS. Besides, we also assume that a user can utilize at most $V^u$ subchannels including licensed and unlicensed ones for the sake of fairness.

First, some notations regarding the subchannel allocation are listed as follows. The subchannel allocation matrix for LTE and D2D users is denoted by
\begin{equation}\label{Allocation_matrix}
A_{(N+M) \times (K + K^u)} = \left(\begin{array}{c} \Phi_{N \times (K + K^u)}\\ \Theta_{M \times (K + K^u)}\end{array}\right),
\end{equation}
where $\Phi_{N \times (K + K^u)} = [\phi_{n,k}]$, and $\Theta_{M \times (K + K^u)} = [\theta_{m,k}]$ stand for the subchannel allocation matrices for the LTE and D2D users, respectively. The values of $\phi_{n,k}$ and $\theta_{m,k}$ are defined as
\begin{equation}\label{D_alpha}
\phi_{n,k}=
\left\{ \begin{gathered}
1,~~\mbox{when subchannel}~\text{SC}_k~\mbox{is allocated to CU}_n,\hfill\\
0,~~\mbox{otherwise},\hfill\\
\end{gathered} \right.
\end{equation}
and
\begin{equation}\label{D_beta}
\theta_{m,k}=
\left\{ \begin{gathered}
1,~~\mbox{when subchannel}~\text{SC}_k~\mbox{is allocated to D}_m,\hfill\\
0,~~\mbox{otherwise}.\hfill\\
\end{gathered} \right.
\end{equation}
Besides, we define the access indicators $s^c_n~(n \in \mathcal{N})$ and $s^d_m~(m \in \mathcal{M})$ to respectively represent whether the LTE and D2D users can access the unlicensed channel. If the LTE user CU$_n$ can access the unlicensed channel, $s^d_n = 1$; otherwise, $s^d_n = 0$. And it is the same for D2D users. We also define $\mathcal{C}_k$ to represent the set of LTE and D2D users to which subchannel SC$_k$ is allocated.

\subsubsection{Interference analysis in the licensed spectrum}
In the licensed subchannels, under the assumption that a subchannel can be allocated to a maximum of one LTE user, the LTE users can only receive the co-channel interference from the underlaid D2D users, while the interference received by D2D users might be from LTE users and other co-channel D2D users. The SINR at the receiver of BS from CU$_n$ over licensed subchannel SC$_k$ can be given by
\begin{equation}\label{SINR_cellular_licensed}
\gamma_{n,k}^c  = \frac{{\phi _{n,k} P^c |h^c_{n,B}|^2 }}{{\sigma ^2  + \sum\limits_{m = 1}^M {\theta _{m,k} P^d |h^d_{m,B}|^2 } }},
\end{equation}
where $h^c_{n,B}$ and $h^d_{m,B}$ represent the channel gains from CU$_n$ and D$^t_m$ to the BS, respectively. The SINR at D$_m^r$ over licensed subchannel SC$_k$ can be expressed as
\begin{equation}\label{SINR_D2D_licensed}
\gamma_{m,k}^d  = \frac{{\theta _{m,k} P^d |h^d_{m,m}|^2 }}{{\sigma ^2  + \sum\limits_{m \ne m',m' = 1}^M {\theta _{m',k} P^d |h^d_{m',m}|^2 }  + \sum\limits_{n = 1}^N {\phi _{n,k} P^c |h^c_{n,m}|^2 } }},
\end{equation}
where $h^d_{m',m}$ and $h^c_{n,m}$ are the channel gains from D$^t_{m'}$ and CU$_n$ to D$^r_{m}$, respectively. The data rates of CU$_n$ and D$_m$ over licensed subchannel SC$_k$ are respectively given by
\begin{equation}\label{rate_licensed}
R_{n,k}^c = B_l \log_2(1 + \gamma_{n,k}^c),~~R_{m,k}^d = B_l \log_2(1 + \gamma_{m,k}^d).
\end{equation}

\subsubsection{Interference analysis in the unlicensed spectrum}
In the unlicensed subchannels, the D2D-U and LTE-U users will not only receive the mutual interference from D2D-U and LTE-U users as in the licensed subchannels, but also the interference from the Wi-Fi users. Therefore, the SINR at the receiver of BS from CU$_n$ over unlicensed subchannel SC$_k$ is
\begin{equation}\label{SINR_cellular_unlicensed}
\gamma _{n,k}^{c,u}  = \frac{{\phi _{n,k} P^c |h^c_{n,B}|^2 }}{{\sigma ^2  + \sum\limits_{m = 1}^M {\theta _{m,k} P^d |h^d_{m,B}|^2 } + I^c_w }},
\end{equation}
where $I^c_w$ is the total interference from Wi-Fi system to BS. Here, the interference can be calculated as $I^c_w = \sum\limits_{q \in \mathcal{Q}} {\frac{{P^w }}{{K^u }}} |h^q_{f,B} |^2$, where $h^q_{f,B}$ is the channel gain from the transmitting Wi-Fi user WU$^q_f$ to the BS. Similarly, the SINR at D$_m^r$ over unlicensed subchannel SC$_k$ can be written as
\begin{equation}\label{SINR_D2D_unlicensed}
\gamma _{m,k}^{d,u}  = \frac{{\theta _{m,k} P^d  |h^d_{m,m}|^2 }}{{\sigma ^2  + \sum\limits_{m \ne m',m' = 1}^M {\theta _{m',k} P^d   |h^d_{m',m}|^2 }  + \sum\limits_{n = 1}^N {\phi _{n,k} P^c  |h^c_{n,m}|^2 } + I^d_w}},
\end{equation}
where $I^d_w$ is the interference from Wi-Fi system to D$^r_m$, whose value is $I^d_w = \sum\limits_{q \in \mathcal{Q}} {\frac{{P^w }}{{K^u }}} |h^q_{f,m}|^2$, with $h^q_{f,m}$ representing the channel gain from the active WU$^q_f$ to D$^r_m$.

The data rates of CU$_n$ and D$_m$ over unlicensed subchannel SC$_k$ are respectively given by
\begin{equation}\label{rate_unlicensed}
R_{n,k}^{c,u} = B_u\log_2(1 + \gamma_{n,k}^{c,u}),~~~R_{m,k}^{d,u} = B_u\log_2(1 + \gamma_{m,k}^{d,u}).
\end{equation}

\section{Duty Cycle Based D2D-U Protocol}%
\label{protocoldesign}

\begin{figure}[!t]
\centering
\includegraphics[width=4.5in]{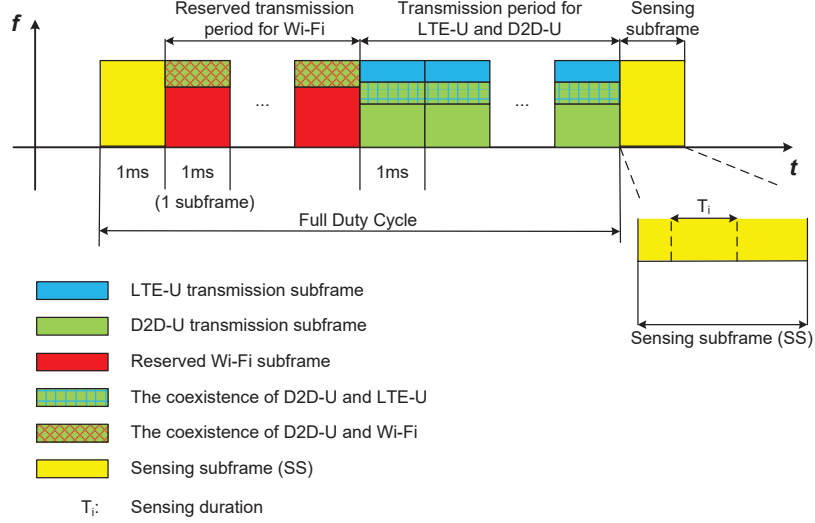}
\vspace{-5mm}
\caption{Duty cycle based protocol for LTE-U and D2D-U users in the unlicensed spectrum.}
\vspace{-7mm}
\label{protocol}
\end{figure}

In this section, we propose a duty cycle \cite{Qualcomm-2014} based protocol for the LTE-U and D2D-U users to share the unlicensed spectrum with Wi-Fi systems. The basic principle of the protocol is to allow the LTE-U and D2D-U users to access the unlicensed spectrum while protecting the incumbent Wi-Fi performance.

\subsection{Overview of the Proposed Protocol}

As illustrated in Fig.~\ref{protocol}, similar to the LTE standard, the timeline is slotted into subframes with length $T_{sub}$ (e.g., 1ms in the LTE standard). There are three types of subframes, namely sensing subframes~(SSs), transmission subframes, and reserved Wi-Fi subframes. The SSs are inserted before the LTE-U and D2D users attempt to select channel and initiate transmission to avoid collision with the ongoing Wi-Fi transmissions. In transmission subframes, the LTE-U and D2D-U users perform transmission as in the conventional LTE standard. Then, to further protect the Wi-Fi performance, we reserve several subframes for Wi-Fi transmission, during which the LTE-U users are not allowed to utilize the unlicensed spectrum.

\subsection{Coexistence Mechanism}

Without modifying current LTE PHY/MAC standards, two mechanism are used to safeguard that LTE-U/D2D-U users do not bring severe interference to their neighboring users in unlicensed spectrum. First, channel selection is performed to choose the cleanest channel avoiding the collision between the ongoing Wi-Fi users and LTE-U/D2D-U users. In the event that no clean channel is available, channel sensing transmission is used to support transmission for D2D-U/LTE-U users.

\subsubsection{Channel selection}

In SS, LTE-U/D2D-U users will scan the unlicensed spectrum and identify the cleanest channel from the $L$ unlicensed channels for the uplink transmission. For an LTE-U/D2D-U user, the transmitter will perform energy detection, and measure the interference level. If the interference is sensed less than the predefined threshold for a sensing duration $T_i$, the channel will be regarded as clean for this LTE-U/D2D-U user. Then, LTE-U/D2D-U users will inform the BS whether they collide with Wi-Fi users according to the measured result. If in the operating channel, the number of interfered users is larger than a given threshold, and there is another cleaner channel available, i.e., the number of interfered users in this channel is less than that in the operating channel, the transmission will be switched to the new channel.

Some technologies are also used to improve detection sensitivity. For example, Wi-Fi preambles are used to estimate the number of neighboring Wi-Fi APs in a given channel. In addition, device-assisted enhancements, such as 802.11k, in which the transmitter sends the request signals and the receiver replies the acknowledgment signal when the request signal is well received, can be used to address the hidden node effect, and thus help to select a better channel.

\subsubsection{Channel sensing transmission}
For most deployments, the channel selection is usually sufficient to meet the coexistence requirements. While in hyper-dense deployment of 5G system, there is a probability that no clean channel can be found. For LTE-U users, carrier-sensing adaptive transmission~(CSAT) algorithm~\cite{Qualcomm-2014} is used to support the coexistence of LTE-U and Wi-Fi users. In the CSAT scheme, LTE-U and Wi-Fi users coexist in a TDM fashion. In particular, a duty cycle is defined where LTE-U users transmit in a fraction of the cycle and gates off in the remaining time to hand over the unlicensed channel to Wi-Fi users.

However, due to the short transmission range and low transmission power of D2D communications, it is possible to share the unlicensed spectrum with Wi-Fi users during the full duty cycle. After the SS, the unlicensed channel can be still utilized by Wi-Fi users to resume the ongoing data transmission. In these reserved subframes, those D2D-U users which have sensed that the channel idle in SS are active in the unlicensed spectrum for data transmission, while other D2D and LTE users are only allowed to utilize licensed subchannels. When the reserved subframes for Wi-Fi transmission expires, all the LTE-U/D2D-U users are activated in the unlicensed subchannels. At the begin of each subframe, the BS allocates licensed and unlicensed subchannels to the LTE/D2D users, in particular, only active LTE-U/D2D-U users have possibility to utilize unlicensed subchannels, which is elaborated in Section \ref{many-to-many}.

\subsection{Analysis of the protocol}

\subsubsection{Compatibility analysis}

For LTE system, unlike the LBT based protocol for D2D-U in~\cite{YWHLSXJ-2016} which requires LBT waveform and transmission modification of current LTE standard, our proposed protocol follows the current LTE PHY/MAC standards, such as frame structures, resource scheduling, and signaling. Thus, it can be directly implemented to current LTE network. And as for Wi-Fi system, D2D-U/LTE-U users also perform energy detection to avoid the collision with Wi-Fi users. Therefore, the LTE-U and D2D-U networks can be a good neighbor of Wi-Fi network.

\subsubsection{Signaling analysis}
To describe the signaling cost over the control channels for the proposed protocol, we assume that $\eta$ messages are required to inform the BS the channel information sensed by a D2D-U/LTE-U user, $\mu$ messages are required for a user to report its location and subchannel estimation results, and $\nu$ messages are needed for the BS to notify a user about the allocated subchannels. In sensing subframe, each D2D-U/LTE-U user $n \in \mathcal{N} \cup \mathcal{M}$ needs to report the sensing result over each channel. Therefore, at most $\eta(M+N)L$  messages are required in the SS. And before each subframe, each LTE-U/D2D-U user needs to report their locations and the subchannel estimation results for subchannel allocation, which requires $\mu(M+N)$ messages. Then, the BS will perform resource allocation process with extra information, and notify each user by sending $\nu(M+N)$ messages.

Note that in one duty cycle, each LTE-U/D2D-U user only performs one energy detection over one channel, thus, the signaling cost is under a tolerable level. In addition, the signaling cost of resource allocation is positively proportional to the number of D2D-U/LTE-U users, which is constrained by the limited subchannel resources. In each subframe, the signaling cost of the resource allocation can be also restricted to a tolerable level. Therefore, the signaling cost of the proposed duty cycle based protocol is acceptable for a practical system.

\section{Problem Formulation}%
\label{problem}

In this section, we first formulate subchannel allocation problem considering both the performance of Wi-Fi and the total sum-rate of LTE and D2D users, and then reformulate this problem into a many-to-many matching problem in consideration of its computation complexity.

\subsection{Sum-rate Maximization Problem Formulation}

Our objective is to maximize the total sum-rate of LTE and D2D users while keeping the interference ranges under a tolerant level by setting the subchannel allocation variables $\left\{\phi_{n,k}, \theta_{m,k}\right\}$ in each subframe.

Since the BS does not hold the information of the interference from LTE-U and D2D-U users to Wi-Fi systems, we use the approximate model described in the Subsection IV-A to evaluate the performance degradation of Wi-Fi system, and add it in the objective function as a penalty term. Assuming that the Wi-Fi users are uniformly located in this plane, the number of interfered Wi-Fi users is therefore positively proportional to the interference range. This can be used as an indicator of the performance degradation of Wi-Fi system. Besides, provided that at least one unlicensed subchannels is allocated to CU$_n$ or D$_m$, the Wi-Fi users in its interference range cannot perform data transmission, and thus, the penalty term is in the form of sign function. Specifically, the penalty items $W^c_{n}$ and $W^d_{m}$ for CU$_n$ and D$_m$ can be respectively given by
\begin{equation}\label{penalty_LTE}
W_{n}^{c} = f_{n}^c~{\mathop{\rm sgn}} \left(\sum\limits_{k \in \mathcal{K}^u }{\phi _{n,k} }\right),~W_{m}^{d} = f_{m}^c~{\mathop{\rm sgn}}\left(\sum\limits_{k \in \mathcal{K}^u }{\phi _{m,k} }\right).
\end{equation}
where ${\mathop{\rm sgn}}(\cdot)$ is the sign function.

Taking the penalty into consideration, the subchannel allocation can be formulated as the following optimization problem:
\begin{subequations}\label{Constraint_silent}
\begin{align}
\max\limits_{\phi_{n,k}, \theta_{m,k}} &\sum\limits_{k \in \mathcal{K}} \sum\limits_{{\mathcal{C}}_k } \left({R _{n,k}^c + R_{m,k}^d}\right)
  + \sum\limits_{k \in \mathcal{K}^u} \sum\limits_{{\mathcal{C}}_k } \left({R_{n,k}^{c,u} + R_{m,k}^{d,u} }\right)
 -\lambda \left(\sum\limits_{n \in \mathcal{N}} W_{n}^c  + \sum\limits_{m \in \mathcal{M}} {W_{m}^d}\right), \label{Objective} \\
s.t.~~&\sum\limits_{n \in \mathcal{N}} {{\phi _{n,k}}}  \leq 1, \forall k \in \mathcal{K}, \label{s_constraint1}\\
&\sum\limits_{k \in \mathcal{K} \cup \mathcal{K}^u} \phi_{n,k} \leq V^u, \forall n \in \mathcal{N}, \sum\limits_{k \in \mathcal{K} \cup \mathcal{K}^u} \theta_{m,k} \leq V^u, \forall m \in \mathcal{M},\label{s_constraint2}\\
&\sum\limits_{n \in \mathcal{N}} \phi_{n,k} + \sum\limits_{m \in \mathcal{M}} \theta_{m,k} \leq V^s, \forall k \in \mathcal{K} \cup \mathcal{K}^u,\label{s_constraint5}\\
&\sum\limits_{k \in \mathcal{K}} \phi_{n,k} \geq 1,\forall n \in \mathcal{N},\sum\limits_{k \in \mathcal{K}} \theta_{m,k} \geq 1,\forall m \in \mathcal{M},\label{s_constraint3}\\
& s^c_n\phi_{n,k} = \phi_{n,k}, s^d_m\theta_{m,k} = \theta_{m,k},\forall k \in \mathcal{K}^u,\label{s_constraint4}
\end{align}
\end{subequations}
where $\lambda \geq 0$ is the sensitivity factor for Wi-Fi systems. Constraint (\ref{s_constraint1}) is given under the assumption that one subchannel can be utilized by at the maximum of one cellular user. Constraints (\ref{s_constraint2}) and (\ref{s_constraint5}) imply that a user utilize at most $V^u$ subchannels, and a subchannel can be allocated to a maximum of $V^s$ users. According to the CA property, each LTE-U or D2D-U user needs to occupy at least one licensed subchannel for control signals, and thus constraint (\ref{s_constraint3}) needs to be satisfied. Constraint (\ref{s_constraint4}) is the sensing constraint, only the LTE-U and D2D-U users which have sensed that the channel is idle can access the unlicensed channel.

Note that the aforementioned problem is a MINLP problem, which is NP-hard \cite{D-1976}. Considering the computational complexity, we reformulate the subchannel allocation as a many-to-many two-sided matching problem, which can be efficiently solved by utilizing the matching games.

\subsection{Matching Formulation}
\label{p_matching}
We consider the set of LTE and D2D users, $\mathcal{U} = \mathcal{M} \cup \mathcal{N}$, and the set of subchannels including licensed and unlicensed, $\mathcal{S} = \mathcal{K} \cup \mathcal{K}^u$, as two disjoint sets of selfish players aiming to maximize their own benefits. Each player can exchange information with one another without extra signaling cost\footnote{The BS is assumed to have the full knowledge of the channel side information (CSI), and performs subchannel allocation based on the obtained CSI.}, that is, the players have complete information about others. In this many-to-many matching model, if subchannel SC$_k$ is assigned to LTE user CU$_n$, then LTE user CU$_n$ is said to be matched with subchannel SC$_k$, and form a matching pair, marked by $(\text{CU}_n,\text{SC}_k)$.

A matching is an assignment of subchannels in $\mathcal{S}$ to users in $\mathcal{U}$, which can be defined as:
\begin{definition}\label{matching}
Given two disjoint sets, $\mathcal{U} = \mathcal{M} \cup \mathcal{N}$ of the users, and $\mathcal{S} = \mathcal{K} \cup \mathcal{K}^u$ of the subchannels, a many-to-many matching $\Psi$ is a mapping from the set $\mathcal{U} \cup \mathcal{S}$ to the set of all subsets of $\mathcal{U} \cup \mathcal{S}$ such that for every user CU$_n \in \mathcal{N}$ or D$_m \in \mathcal{M}$, and subchannel SC$_k \in \mathcal{S}$:
\begin{enumerate}[(1)]
\item $\Psi(\text{CU}_n) \in \mathcal{S}$, $\Psi(\text{D}_m) \in \mathcal{S}$;\label{c1}
\item $\Psi(\text{SC}_k) \in \mathcal{U}$;\label{c2}
\item $|\Psi(\text{CU}_n)| \leq V^u, |\Psi(\text{D}_m)| \leq V^u$;\label{c3}
\item $|\Psi(\text{SC}_k)| \leq V^s$;\label{c7}
\item $|\Psi(\text{CU}_n) \cap \mathcal{K}| \geq 1, |\Psi(\text{D}_m) \cap \mathcal{K}| \geq 1$;\label{c4}
\item $s^d_m = 0 \Leftrightarrow \Psi(\text{D}_m) \cap \mathcal{K}^u = \emptyset$, $s^c_n = 0 \Leftrightarrow \Psi(\text{CU}_n) \cap \mathcal{K}^u = \emptyset$;\label{c5}
\item $\text{SC}_k \in \Psi(\text{CU}_n) \Leftrightarrow \text{CU}_n \in \Psi(\text{SC}_k)$, $\text{SC}_k \in \Psi(\text{D}_m) \Leftrightarrow \text{D}_m \in \Psi(\text{SC}_k)$.\label{c6}
\end{enumerate}
\end{definition}

Conditions (\ref{c1}) and (\ref{c2}) state that each LTE or D2D user is matched with a subset of subchannels, and each subchannel is matched with a subset of users. Conditions (\ref{c3}) and (\ref{c7}) show the utilization constraints for a user and a subchannel. Due to the CA requirement, the users need to occupy at least one licensed subchannel, as expressed in condition (\ref{c4}). Condition (\ref{c5}) implies that only those users sensed idle unlicensed subchannel can utilize the unlicensed subchannels.

Considering mutual interference items in (\ref{SINR_D2D_licensed}) and (\ref{SINR_D2D_unlicensed}), any D2D user's sum-rate over its allocated subchannels SC$_k$ is related to the set of other LTE users and D2D pairs sharing this subchannel. Besides, the penalty term in (\ref{Objective}) indicates that the objective of the LTE-U and D2D users is relevant to other users operating in the unlicensed spectrum as well. Thus, each user cares about not only which subchannel it is matched with, but also the set of users matching with the same subchannel. For this reason, the aforementioned matching game is a many-to-many matching game with \emph{externalities}~\cite{AM-1992} or \emph{peer effects} \cite{ECABA-2011}.

Affected by the peer effects, the outcome of this matching game greatly depends on the dynamic interactions among the users sharing the subchannels. To better describe the selection behavior and decision process of each player, we introduce a concept of preference relation $>$ for both users and subchannels. For any two subchannels $\text{SC}_k, \text{SC}_{k'} \in \mathcal{S}, k \ne k'$, and any two matchings $\Psi, \Psi', \text{SC}_k \in \Psi(\text{D}_m), \text{SC}_{k'} \in \Psi'(\text{D}_m)$:
\begin{equation}\label{preference1}
(\text{SC}_k,\Psi) >_{D_m} (\text{SC}_{k'} ,\Psi') \Leftrightarrow R_{m,k}^d (\Psi) > R_{m,k'}^d (\Psi'),
\end{equation}
where $R_{m,k}^{d}$ is related to the current subchannel allocation results. If D2D user D$_m$ has not been allocated to unlicensed subchannels, the data rate $R_{m,k}^{d}$ needs to deduct the penalty item. This implies that the D2D user D$_m$ prefers SC$_k$ in $\Psi$ to SC$_{k'}$ in $\Psi'$ if D$_m$ can have a higher data rate over SC$_k$ than SC$_{k'}$. The same process will be done for an LTE user CU$_n \in \mathcal{N}$. LTE-U user CU$_n$ will also prefer the subchannel which can achieve higher data rate.

As for any subchannel SC$_k \in \mathcal{S}$, its preference relation $>_{\text{SC}_k}$ over the set of users can be given in an uniform method. For any two subsets of users $T, T¡¯ \in \mathcal{U}, T \ne T'$, and any two matchings $\Psi,\Psi', T = \Psi(\text{SC}_k), T' = \Psi'(\text{SC}_k)$:
\begin{equation}\label{preference4}
(T,\Psi) > _{\text{SC}_k} (T' ,\Psi') \Leftrightarrow R_{\text{SC}_k} (\Psi)  > R_{\text{SC}_{k'}} (\Psi'),
\end{equation}
where $R_{\text{SC}_k}$ also includes the penalty items. This indicates that subchannel SC$_k$ prefers the set of users $T$ to $T'$ only when SC$_k$ can get a higher data rate from $T$. $(T,\Psi) \geq _{\text{SC}_k} (T' ,\Psi')$ is also used to indicate that subchannel SC$_k$ likes the set of users $T$ at least as well as $T'$.

Different from traditional many-to-many matchings in which the players' preferences are substitutable, subchannels' preferences do not satisfy \emph{substitutability}. Specifically, given a subchannel SC$_k \in K$, let $\mathcal{T}_k \subseteq \mathcal{U}$ represent its most preferred user set  that containing two D2D pairs D$_m$ and D$_i$. Besides, the data rate $R_{m,k}^d$ of D$_m$ is higher than $R_{i,k}^d$ of D$_i$ when they utilize subchannel SC$_k$ independently. If D$_m \notin \mathcal{T}_k$, then it is not necessary that D$_i \in \mathcal{T}_k/\{\text{D}_m\}$. Due to the mutual interference, the data rate may have changed after D$_m$ is removed from $\mathcal{T}_k$, and thus, SC$_k$ may not prefer D$_i$ any more.

Due to the externalities, the many-to-many matching model in this work is more complicated than the conventional two-sided matching models. Under traditional definition of stable matching\footnote{Traditional stable matching refers to a matching in which there do not exist two players from opposite sets prefer each other to at least one of their current matches such that they form a new matching pair together for the sake of their interests, that is, there are no blocking pairs in a stable matching.} in \cite{AM-1992}, there is no guarantee that a stable matching exists even in many-to-one matchings. Because of the lack of substitutability, traditional deferred acceptance algorithm \cite{AM-1992} cannot be applied to this model any more. To solve this matching problem, we introduce the swap matching~\cite{ECABA-2011} and propose a matching algorithm in Section~\ref{many-to-many}.

\section{Many-to-Many Matching-Based Subchannel Allocation}%
\label{many-to-many}

In this section, we propose a matching algorithm to solve the problem formulated in Section \ref{p_matching}. We first introduce the notations and definitions of \emph{swap matching} and \emph{stability} into our many-to-many matching model, and then elaborate on the swap matching algorithm.

\subsection{Notations and Definitions}

The concepts of \emph{swap matching} and \emph{swap-blocking pair} are defined as below.

\begin{definition}\label{swap_matching}
Given a matching $\Psi$, two matching pairs $(\text{U}_i, \text{SC}_p)$ and $(\text{U}_j,\text{SC}_q)$ with SC$_p \in \Psi(\text{U}_i)$, SC$_q \in \Psi(\text{U}_j)$, SC$_p \notin \Psi(\text{U}_j)$, and SC$_q \notin \Psi(\text{U}_i)$,
\textbf{a swap matching} is defined as:
\begin{equation}\label{swap}
\Psi_{i,p}^{j,q} =  \{ \Psi \backslash \{(\text{U}_i, \text{SC}_p),(\text{U}_j,\text{SC}_q)\}\} \cup \{(\text{U}_i, \text{SC}_q),(\text{U}_j,\text{SC}_p)\}.
\end{equation}
\end{definition}

A swap matching is generated via swap operations, which is the two-sided version of the \emph{exchange} operation \cite{KD-2005,R-2008}. In the swap operation, a pair of players exchange their matches while all other matchings remain unchanged. Different from conventional strategy change in one-to-one matching performed by the individual player, the swap operation needs to be approved by both involved players. In the following, we provide the conditions in which the swap operations can be approved by introducing the concepts of \emph{swappable set} and \emph{swap-blocking pair}.

\begin{definition}\label{swappable}
For LTE-U user CU$_n$ or D2D user D$_m$, its \textbf{swappable set} is defined as a subchannel subset in which the user can swap for subchannel via swap matching. Specifically, if the sensing vector $s_m^d = 1$, the swappable set $\Omega_m$ of D$_m$ is subchannels set $\mathcal{S}$ including licensed and unlicensed ones; otherwise, its swappable set $\Omega_m$ is licensed subchannels set $\mathcal{K}$. And it does the same for LTE user CU$_n$.
\end{definition}

Note that only those users which have sensed that the channel is idle can access the unlicensed subchannel, the users failed to sense idle unlicensed subchannels can only swap for licensed subchannels in the swappable sets.

\begin{definition}\label{blocking_pair}
Provided a matching $\Psi$ and a pair $(\text{U}_i,\text{U}_j), \text{U}_i, \text{U}_j \in \mathcal{U}$, $\text{U}_i$ and $\text{U}_j$ are matched in $\Psi$, and let $\Omega_i$ and $\Omega_j$ respectively represent the swappable sets of $\text{U}_i$ and $\text{U}_j$. If there exist subchannels SC$_p \in \Psi(\text{U}_i), \text{SC}_p \in \Omega_j$, SC$_q \in \Psi(\text{U}_j)$, and SC$_q \in \Omega_i$ such that:
\begin{enumerate}[(1)]
\item $\forall t \in (\text{U}_i, \text{U}_j, \text{SC}_p, \text{SC}_q), (\Psi_{i,p}^{j,q},\Psi_{i,p}^{j,q}(t)) \geq_{t} (\Psi,\Psi(t))$,
\item $\exists t \in (\text{U}_i, \text{U}_j, \text{SC}_p, \text{SC}_q), (\Psi_{i,p}^{j,q},\Psi_{i,p}^{j,q}(t)) >_{t} (\Psi,\Psi(t))$,
\item $\forall t \in (\text{U}_i, \text{U}_j), |\Psi_{i,p}^{j,q}(t) \cap \mathcal{K}| \geq 1$,
\end{enumerate}
then the swap matching $\Psi_{i,p}^{j,q}$ is \textbf{approved}, and the pair $(\text{U}_i,\text{U}_j)$ is called \textbf{a swap-blocking pair} in the matching $\Psi$.
\end{definition}

The third condition in \textbf{Definition \ref{blocking_pair}} is to satisfy the CA requirement in which each user needs to utilize at least one licensed subchannel. The definition implies that once a swap matching is approved, at least one player's data rate will increase, which leads to the increase in the total data rate.

\begin{definition}\label{stable}
A matching $\Psi$ is \textbf{two-sided exchange-stable (2ES)} if and only if there does not exist a swap-blocking pair.
\end{definition}

Intuitively speaking, from the perspective of network, a matching $\Psi$ is said to be \textbf{2ES} implies that there is not any user $U_i$ or subchannel SC$_q$, in which $U_i$ prefers another subchannel SC$_p$ to its match SC$_q$, or SC$_q$ likes another user $U_j$ rather than its match $U_i$. Such a network-wide stable can be achieved by guaranteeing the involved players are beneficial from the swap operations, given the externalities in current matching $\Psi$.

\subsection{Algorithm Description}

With the notations of swap matching and the definition of stability, we propose a user-subchannel matching algorithm (Algorithm \ref{Sub_Matching}) to obtain a \textbf{2ES} matching. This algorithm is a extension of the many-to-one matching algorithm proposed in \cite{FMWSM-2013} with constraints that $|\Psi(\text{CU}_n)|~\geq~V^u$, $|\Psi(\text{D}_m)| \geq V^u$, and $|\Psi(\text{SC}_k)| \geq V^s$.

As a part of Algorithm \ref{Sub_Matching}, each LTE user or D2D user needs to maintain a preference list. The preference list is established according to the following principles:
\begin{enumerate}[(1)]
\item The subchannels in the preference list need to be contained in the swappable set.
\item The matched subchannel is removed from the preference list for each user.
\item The subchannels which have matched with $V^s$ users is removed from the preference list.
\item If the user is unmatched, i.e., the user has not been allocated to any subchannels, the licensed subchannels have priorities over the unlicensed ones.
\item The preference list is established based on the data rate over each subchannel.
\end{enumerate}

In Algorithm \ref{Sub_Matching}, each user will send a proposal to the BS. According to definition \ref{swappable}, the proposed subchannel needs to be contained by swappable set. For each user, removing the matched subchannel is to avoid multiple proposals for the same subchannel. In addition, under the utilization constraints for a subchannel, the users can only send proposal to the available subchannels. The forth principle is designed in accordance with the CA requirement. This implies that if the user cannot compete for a licensed subchannel, the user needs to be silent. And according to the definition of preference relation in (\ref{preference1}), the preference list is maintained based on the data rate. Due to the externalities, the preference list is dynamic in the swap matching process. Thus, in each iteration, the preference list will be updated based on the current matching.

The key idea of Algorithm \ref{Sub_Matching} is to consider approving swap matchings among the players so as to obtain a \textbf{2ES} matching. The algorithm is composed of two phases: initialization phase and swap matching phase. In the initialization phase, the BS will evaluate the channel gains for all users and interference from WiFi system. The swap matching phase contains multiple iterations in which the BS keeps executing the swap matching if there exist swap-blocking pairs, and updates the current matching. Note that the higher a user's data rate is, the higher probability it has to be accepted by the subchannel. In each iteration, the user $U_i$ updates its preference list, and sends a proposal to the subchannel SC$_q$ ranked the first in the preference list unless it has been matched with $V^u$ subchannels. The acceptance can be regarded as a swap operation $\Psi_{i,0}^{0,q}$, where the element $\{0\}$ denotes a virtual user or subchannel. If this swap matching is approved, the proposed user $U_i$ is accepted by the subchannel SC$_q$, and the matching is updated. Then, the BS will search other swap-blocking pairs and execute the swap matching to renew the current matching. The iterations stop until current matching is the same as the matching in the last iteration, and a final matching is determined.

\begin{algorithm}[!t]
\caption{User-Subchannel Matching Algorithm for LTE Users and D2D Pairs.}\label{Sub_Matching}
\KwIn{Set of users $\mathcal{U}$; set of subchannels $\mathcal{S}$; sensing vectors $s^c_n, s^d_m$.}\vspace{-1.5mm}
\KwOut{A 2ES matching $\Psi$.}\vspace{-1.5mm}
\Begin
{\vspace{-1.5mm}
\While{$\Psi \ne \Psi'$}
{\vspace{-1.5mm}
    Let $\Psi' = \Psi$\;\vspace{-1.5mm}
    The preference lists are updated based on the current matching $\Psi$\;\vspace{-1.5mm}
    \eIf{$\text{U}_i$ has not been matched with $V^u$ subchannels}
    {\vspace{-1.5mm}
    $\text{U}_i$ sends a proposal to the first subchannel SC$_q$ in the preference list\; \vspace{-1.5mm}
    \eIf{swap operation $\Psi_{i,0}^{0,q}$ is approved}
    {\vspace{-1.5mm}
    User $\text{U}_i$ matches with subchannel SC$_q$\;\vspace{-1.5mm}
    The current matching $\Psi$ is replaced by swap matching $\Psi_{i,0}^{0,q}$\;\vspace{-1.5mm}
    }
    {\vspace{-1.5mm}
    User $U_i$ cannot get access to SC$_q$\;\vspace{-1.5mm}
    }\vspace{-1.5mm}
    The BS searches for swap-blocking pairs\;\vspace{-1.5mm}
    \eIf{swap operation $\Psi_{i,p}^{j,q}$ is approved}
    {\vspace{-1.5mm}
    User $U_i$ exchanges its match SC$_p$ with $\text{U}_j$ for subchannel SC$_q$\;\vspace{-1.5mm}
    The current matching $\Psi$ is replaced by swap matching $\Psi_{i,p}^{j,q}$\;\vspace{-1.5mm}
    }
    {\vspace{-1.5mm}
    User $\text{U}_i$ keeps its match SC$_p$\;\vspace{-1.5mm}
    }\vspace{-1.5mm}
    }
    {\vspace{-1.5mm}
    User $\text{U}_i$ keeps its matches.
    }\vspace{-1.5mm}
}\vspace{-1.5mm}
Terminate with the final matching result $\Psi$\;\vspace{-1.5mm}
}
\end{algorithm}

\section{Performance Analysis}
\label{Discussion}

In this section, we analyze the effectiveness and efficiency of the proposed algorithm, and remark some key properties of the LTE-U/D2D network. In the first part, the effectiveness and efficiency of the proposed algorithm is proved. Then, we discuss how the sensitivity factor $\lambda$ impacts the subchannel allocation strategy.


\subsection{Stability, Convergence, Complexity, and Optimality}
\label{property}

Given the proposed Algorithm \ref{Sub_Matching}, we give remarks on the stability, convergence, complexity, and optimality.

\subsubsection{Stability and Convergence}
We first provide the stability and convergence of Algorithm \ref{Sub_Matching}.

\begin{lemma}\label{covergency}
Phase II in Algorithm \ref{Sub_Matching} converges after a limited number of swap operations.
\end{lemma}

\begin{IEEEproof}
In each iteration of Algorithm \ref{Sub_Matching}, the matching $\Psi$ is updated after a swap operation. Without loss of generality, we assume that after swap operation $l$, the matching result is updated by swap matching $\Psi_l = {\Psi_{l-1}}_{i,p}^{j,q}$. According to definition \ref{blocking_pair}, after swap operation $l$, the sum-rates of subchannel SC$_p$ and SC$_q$ satisfy $R_{\text{SC}_p}(\Psi_l) \geq R_{\text{SC}_p}(\Psi_{l - 1})$ and $R_{\text{SC}_q}(\Psi_l) \geq R_{\text{SC}_q}(\Psi_{l - 1})$, and these two equations cannot hold at the same time, while the sum-rates of other subchannels remain unchanged. Therefore, the total sum-rate over all subchannels strictly increases.

Note that the number of potential swap-blocking pairs is finite since the number of users is limited, and the total sum-rate has an upper bound due to limited subchannels. Therefore, there exists a swap operation after which no swap-blocking pairs can be found and the total sum-rate stops increasing. Then Algorithm \ref{Sub_Matching} converges.
\end{IEEEproof}

\begin{proposition}\label{stablity}
Upon the convergence of Phase II, Algorithm \ref{Sub_Matching} reaches a \textbf{2ES} matching.
\end{proposition}

\begin{IEEEproof}
The proof follows from these two considerations. First, the swap operations only occur when the players' data rate strictly increases. Second, due to the convergency of Phase II, for any user $\text{U}_i \in \mathcal{U}$, it cannot find another user $\text{U}_j \in \mathcal{U}$ to form a swap-blocking pair with their matches when Algorithm \ref{Sub_Matching} terminates. The matches of user $M_i$ must be the best choice in current matching. Hence, the terminal matching obtained by Algorithm \ref{Sub_Matching} is \textbf{2ES}.
\end{IEEEproof}

\subsubsection{Complexity}
Having proved the convergence of Algorithm \ref{Sub_Matching}, we then discuss its computational complexity.

Note that in the swap matching phase, a number of iterations are performed to reach the \textbf{2ES} matching. In every iteration, the BS needs to search for swap-blocking pairs and all the approved swap operations are executed. Thus, the complexity of the swap matching phase lies in the number of both iterations and potential swap matchings in each iteration.

\begin{proposition}\label{complexity}
In the $t$-th iteration of Algorithm \ref{Sub_Matching}, at most $(M+N)*[(M+N-1)*(t-1) + 1]$ swap matchings need to be considered.
\end{proposition}
\begin{IEEEproof}
In each iteration of Algorithm \ref{Sub_Matching}, at most $M + N$ users send proposal to the subchannels which rank first in their preference lists. Therefore, in this step, at most $M + N$ swap matchings need to be considered.

If the proposals from users are accepted by subchannels, they might execute swap matchings with the existing matches. For user $U_i$, it sends proposal to subchannel SC$_p$ and is accepted. According to definition \ref{swap_matching}, this match can only execute swap matchings with matches which do not contain $\text{U}_i$ and SC$_p$. In each iteration, at most 1 match can be added to the current matching for each user. Therefore, for match pair $(\text{U}_i,\text{SC}_q)$, there are at most $(M+N-1)*(t-1)$ potential swap matchings in the $t$-th iteration. In the worst case, all the proposal for users are accepted by subchannels, and thus, there are a maximum of $(M+N)*(M+N-1)*(t-1)$ potential swap matchings.

Above all, at most $(M+N)*[(M+N-1)*(t-1) + 1]$ swap matchings need to be considered in the $t$-th iteration. In practice, one iteration requires a significantly low number of swap operations, since only a small number of proposals from users can be accepted.
\end{IEEEproof}

\subsubsection{Optimality}

We show whether Algorithm \ref{Sub_Matching} can achieve an optimal matching.

\begin{proposition}\label{local_maxima}
All local maxima of total sum-rate corresponds to a \textbf{2ES} matching.
\end{proposition}
\begin{IEEEproof}
Suppose the total sum-rate of matching $\Psi$ is a local maximum value. If $\Psi$ is not a \textbf{2ES} matching, then there exists at least one swap-blocking pair, and any swap matching strictly increases data rates according to definition \ref{swap_matching}. However, this is in contradiction to the assumption that $\Psi$ is a local maximum value. Therefore, $\Psi$ must be \textbf{2ES}.
\end{IEEEproof}

However, not all \textbf{2ES} matchings obtained from Algorithm \ref{Sub_Matching} are local maxima of total data rates. For example, there exists possibility that a user $\text{U}_i$ does not approve a swap matching $\Psi_{i,p}^{j,q}$ since its data rate will decrease, but the other user $\text{U}_j$ will benefit from this swap matching, and the sum-rates of SC$_p$ and SC$_q$ increase as well. The total sum-rates will increase at the expense of stability if the swap operation is forced to execute.

To obtain a global optimum matching, we utilize a algorithm (GO Algorithm) proposed in~\cite{ECABA-2011} by utilizing a Markov chain Monte Carlo heat bath method. In GO Algorithm, the swap matching does not need to be approved any more, instead, a swap matching $\Psi_{i,p}^{j,q}$ is executed
with a probability $P_{swap}$ which depends on the total sum-rate as shown below:
\begin{equation}\label{probability}
P_{swap} = \frac{1}{1+e^{-T[R_{total}(\Psi_{i,p}^{j,q})-R_{total}(\Psi)]}},
\end{equation}
where $T$ is a probability parameter. The algorithm keeps track of the optimum matching found so far, even when it moves to worse matchings. After sufficiently large amount of iterations, the matching moves towards the global optimal one \cite{O-2001}.

\subsection{Selection of the Sensitivity Factor $\lambda$}
\label{selection}

Let $R_{max}^d$ be the maximum rate for a D2D-U user over an unlicensed subchannel, $R_{max}^c$ be the maximum rate for an LTE-U user over one unlicensed subchannel, and generally $R_{max}^d/L_d \geq R_{max}^c/L_c$. How the value of the sensitivity factor $\lambda$ tunes the performance can be analyzed in the following cases.

\subsubsection{$\lambda \geq R_{max}^d/(2L_d)$}

This case implies that neither D2D nor LTE users can get access to an unlicensed subchannel, the value of penalty terms is sufficiently large that cannot satisfy all the conditions of swap matching. Therefore, in this case, any D2D or LTE users cannot utilize unlicensed subchannels, the LTE and D2D users can only utilize the licensed spectrum for total sum-rate maximization.

\subsubsection{$R_{max}^c/(2L_c) \leq \lambda \leq R_{max}^d/(2L_d)$}

This case implies that D2D users can get access to unlicensed subchannels. In this case, any LTE users cannot utilize unlicensed subchannels. As for the D2D users, they can utilize both licensed and unlicensed subchannels. If one D2D user get access to a unlicensed subchannel, those D2D users whose interference ranges overlap with the accessed one will become more easier to get access to the unlicensed subchannels because the increased interference ranges will be less than that when this D2D user is the first one to get access to the unlicensed subchannels. Thus, in the view of geography, those D2D users allocated to unlicensed subchannels trend to form several clusters.

\subsubsection{$0 < \lambda < R_{max}^c/(2L_c)$}

This case means that both LTE and D2D users can get access to unlicensed subchannels. Similar to case 2, the users accessed to unlicensed subchannels also form several clusters. In addition, the accessed LTE users will decrease with the value of $\lambda$ grows because of the large interference range, that is, more unlicensed subchannels are allocated to D2D users and the LTE users are allocated to more licensed subchannels.

\subsubsection{$\lambda = 0$}

This case is the same as resource allocation problem in licensed scenario. In this case, both licensed and unlicensed subchannels are uniform, D2D and LTE users will make full use of the whole spectrum to maximize the total sum-rate.

\section{Simulation Results}%
\label{Simulation}

\begin{table}
\centering
\caption{Parameters for Simulation} \label{parameters}
\vspace{-3mm}
\begin{tabular}{|p{5cm}|p{4cm}|}
 \hline \textbf{LTE-U and D2D-U Parameters} & \textbf{Values}\\
 \hline
  Cell radius & 500 m\\
  D2D communication radius $D_{max}$ & 20 m \\
  LTE's transmit power $P^c$ & 17 dBm\\
  D2D's transmit power $P^d$ & 10 dBm \\
  Subchannel bandwidth $B_l$& 180 kHz \\
  Number of subchannels & 10\\
  Carrier frequency & 1.9 GHz  \\
  Noise figure & 5 dB \\
  Decay factor of the path loss $\alpha$ & 2.2\\
  Power gains factor $G$ & -33.58 dB\\
  Shadow fading standard deviation & 4 dB\\
  Maximum number of subchannels $V^s$ & 4\\
  Maximum number of users $V^u$ & 4\\
 \hline \hline \textbf{Wi-Fi Parameters} & \textbf{Values}\\
 \hline
  Number of subchannels & 20\\
  Subchannel bandwidth $B_u$& 180 kHz \\
  Wi-Fi user's transmit power $P^w$ & 23 dBm\\
  Number of APs $Q$ & 3\\
  LTE-U interference radius $L_c$ & 50 m \\
  LTE-U interference radius $L_d$ & 23 m \\
\hline
\end{tabular}
\end{table}

In this section, we present the simulation results of Algorithm \ref{Sub_Matching}, in comparison to the GO Algorithm, a greedy algorithm, and the scenario without D2D, where all the users are LTE ones. In the greedy algorithm, the users will maintain a static preference list, and send proposal to the subchannels according to the preference list. We set the number of iterations as $10^6$, $T = 0.5$ in GO Algorithm such that the outcome of GO Algorithm can be regarded as the upper bound of the data sum-rate. Note that the upper bound is unrealistic since the computational complexity is rather high. And the subchannels in the scenario without D2D are also allocated by Algorithm \ref{Sub_Matching}. In this simulation, we consider a single cell layout, where the LTE and D2D users are distributed randomly, and the communication distance of D2D users cannot exceed a predefined value. The simulation parameters based on existing LTE-Advanced specifications \cite{3GPP-2014} are given in Table \ref{parameters}. Note that the transmission power of Wi-Fi user is over the whole unlicensed channel, while the transmission power of LTE or D2D user is over one subchannel.

\begin{figure}[!t]
\centering
\includegraphics[width=3.5in]{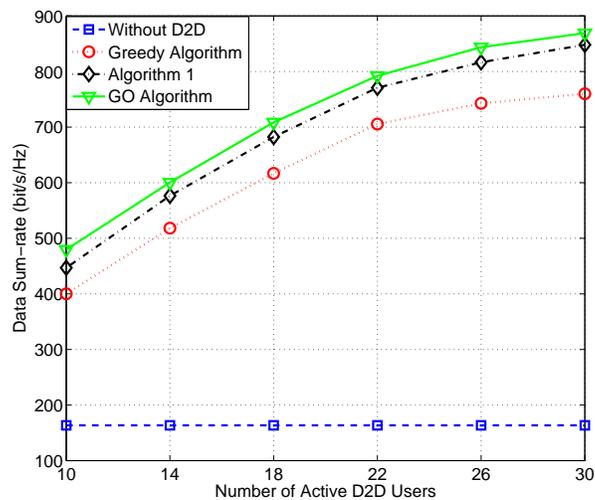}
\vspace{-5mm}
\caption{Sum-rate vs. number of active D2D users, with the number of LTE users $M = 10$ and sensitivity factor $\lambda = 0.1$.}
\vspace{-7mm}
\label{performance}
\end{figure}

Fig. \ref{performance} shows the data sum-rate vs. the number of active D2D users $N$, with the number of LTE users $M = 10$ and the sensitivity factor $\lambda = 0.1$. We observe that the sum-rate increases with $N$. This is because the number of concurrent transmission links grows while any two links which bring severe mutual interference are not allowed to exist in the same subchannel. However, it also shows that the sum-rate becomes saturated when $N > 26$, as the number of subchannels is not sufficient to support more D2D users. In addition, it can be observed that the sum-rate obtained by Algorithm \ref{Sub_Matching} is 10.6\% higher than the greedy algorithm, and 317.3\% higher than the scenario without D2D, while it only has 3.7\% gap to the upper bound when $N = 22$. This further implies that the BS can make full use of the unlicensed spectrum resources via D2D communications. The simulation results correspond to analysis in Section \ref{property}.

\begin{figure}[!t]
\centering
\includegraphics[width=3.5in]{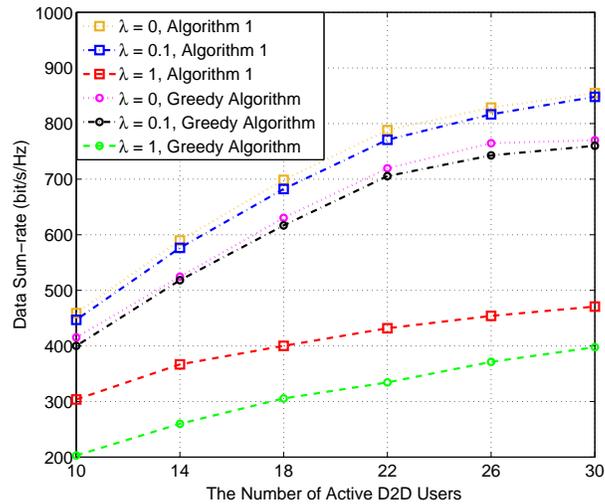}
\vspace{-5mm}
\caption{Sum-rate vs. number of active D2D users, with the number of LTE users $M = 10$. }
\vspace{-7mm}
\label{sumrate_lambda}
\end{figure}

Fig.~\ref{sumrate_lambda} shows the sum-rate v.s. the number of active D2D users with the number of LTE users $M = 10$. It can be easily observed that the total sum-rate obtained by the same algorithm will decrease as the sensitivity factor $\lambda$ increases. According to the discussions in Section \ref{selection}, $\lambda = 0$ means D2D and LTE users can use unlicensed spectrum. Since the subchannels are sufficient, the system sum-rate is the maximum. $\lambda = 1$ means that almost all LTE and D2D users cannot use unlicensed spectrum, and thus the system sum-rate is the minimum. Because of less available subchannels, the data sum-rate decreases as the sensitivity factor $\lambda$ increases. In addition, we can also find out that the data sum-rate obtained by Algorithm \ref{Sub_Matching} is always higher than that obtained by the greedy algorithm with the same sensitivity factor $\lambda$. In particular, the data sum-rate is 100 bit/s/Hz higher when $\lambda = 1$, which implies that Algorithm \ref{Sub_Matching} can utilize the spectrum more efficiently.

\begin{figure}[!t]
\centering
\includegraphics[width=3.5in]{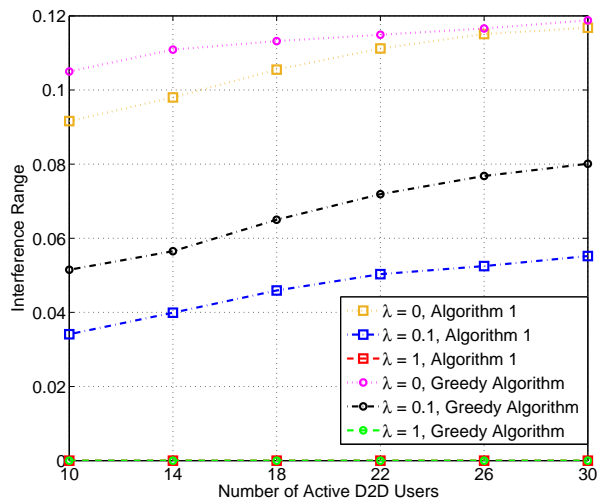}
\vspace{-5mm}
\caption{Interference ranges vs. number of active D2D users, with the number of LTE users $M = 10$.}
\vspace{-7mm}
\label{interfering_lambda}
\end{figure}

In Fig.~\ref{interfering_lambda}, we provide the interference ranges vs. the number of active D2D users $N$ with the number of LTE users $M = 10$. We use a uniform sampling and judge whether this sampling point is in the interference range of any LTE or D2D user using the unlicensed spectrum. The percentage of sampling points in the interference ranges is regarded as the interference ranges. From Fig.~\ref{interfering_lambda}, it can be observed that the interference ranges obtained by the same algorithm will decrease as the sensitivity factor $\lambda$ grows. However, we can find out that the decrease in interference ranges is at the expense of the data sum-rate from Fig.~\ref{sumrate_lambda}. Therefore, we can utilize the unlicensed spectrum according to different traffic requirements by properly setting the value of $\lambda$. In addition, we can also observe the interference range obtained by Algorithm \ref{Sub_Matching} is lower than that obtained by the greedy algorithm with the same $\lambda$, except $\lambda = 1$, where LTE and D2D users cannot utilize the unlicensed spectrum. This implies that Algorithm \ref{Sub_Matching} outperforms the greedy algorithm not only on the data sum-rate, but also on the interference ranges.

\begin{figure}[!t]
\centering
\includegraphics[width=3.5in]{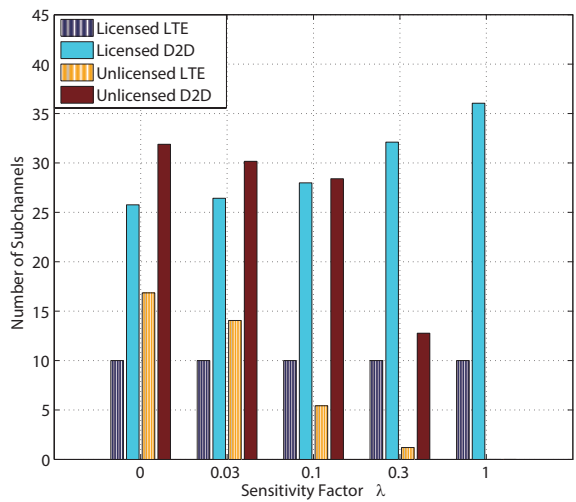}
\vspace{-5mm}
\caption{Number of subchannels vs. sensitivity factor $\lambda$, with the number of LTE users $M = 10$ and the number of D2D Pairs $N = 15$.}
\vspace{-7mm}
\label{subchannel_lambda}
\end{figure}

Fig.~\ref{subchannel_lambda} provides that the number of subchannels in both licensed and unlicensed spectra vs. the sensitivity factor~$\lambda$ with the number of LTE users $M = 10$ and the number of D2D pairs $N = 15$. Note that each LTE user needs to occupy at least one licensed subchannel, and two LTE users are not allowed to transmit on the same subchannel. Thus, each LTE user will only utilize one licensed subchannel. Based on the constraint that a user can utilize at most $V^u = 4$ subchannels, the total subchannels of D2D users cannot exceed 60 subchannels, and the total subchannels for LTE users cannot exceed 40 subchannels. It can be also observed that when the value of $\lambda$ increases, the unlicensed subchannels for LTE and D2D users will decrease for the protection of Wi-Fi system. However, due to the smaller interference ranges of D2D communications, the proportional reduction in unlicensed subchannels for D2D users is lower than that for LTE users. This is consistent with the discussions in Section \ref{selection}.

\begin{figure}[!t]
\centering
\includegraphics[width=3.5in]{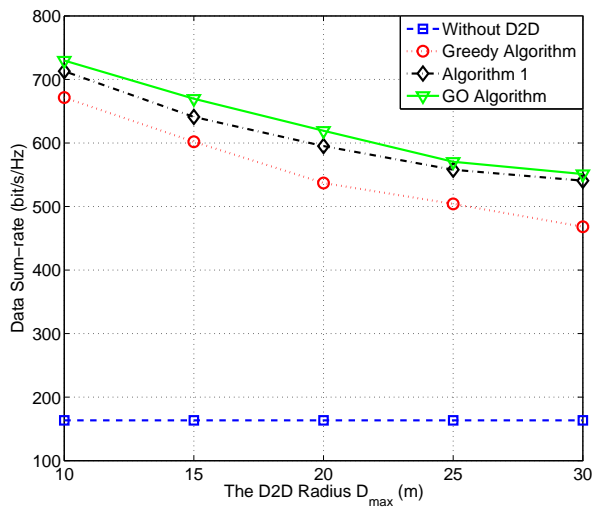}
\vspace{-5mm}
\caption{Data sum-rate vs. D2D radius $D_{max}$, with the number of LTE users $M = 10$, the number of D2D users $N = 20$, and sensitivity factor $\lambda = 0.1$.}
\vspace{-7mm}
\label{D2D_radius}
\end{figure}

Fig.~\ref{D2D_radius} shows the data sum-rate vs. the D2D radius~$D_{max}$ with the number of LTE users $M = 10$, the number of D2D users $N = 20$, and the sensitivity factor $\lambda = 0.1$. In Fig.~\ref{D2D_radius}, we can learn that the data sum-rate will decrease as the D2D radius $D_{max}$ grows. This is because the received power of D2D users downgrades as the transmission radius of D2D users upgrades. It can be also observed that Algorithm \ref{Sub_Matching} always outperforms the greedy algorithm, and approaches to the upper bound with different values of D2D radius.

\section{Conclusions}%
\label{Conclusion}

In this paper, we investigate the D2D-U technology, in which the D2D users operate as an underlay to the LTE system in both licensed and unlicensed spectra. A duty cycle based protocol is designed for LTE-U and D2D-U users while protecting the existing Wi-Fi systems. Considering the complicated mutual interference between LTE, D2D, and Wi-Fi systems, we study the subchannel allocation problem for D2D and LTE users sharing both licensed and unlicensed subchannels to leverage the performance degradation in Wi-Fi systems and the maximization of the sum-rate in LTE/D2D networks. Specifically, we formulate the allocation problem as a many-to-many matching game with externalities, and develop a low-complexity user-subchannel swap matching algorithm. In addition, power control can be done in parallel with subchannel assignment. Analytical and simulation results show that enabling D2D-U communications can significantly improve the system sum-rate. Besides, the subchannel allocation strategy for LTE-U and D2D users is closely related to how the BS adjusts the interference to Wi-Fi systems. In an aggressive strategy where the Wi-Fi performance degradation is not considered seriously, the BS allows more D2D and LTE users to transmit on the unlicensed spectrum. On contrary, in a Wi-Fi friendly strategy, the BS tends to permit only a small fraction of D2D users to transmit on the unlicensed spectrum.

\end{document}